\documentclass[conference]{IEEEtran}
\usepackage{multirow}
\usepackage{array}
\usepackage{makecell}
\usepackage{float}
\usepackage{booktabs}
\usepackage{hyperref}
\hypersetup{hidelinks}
\usepackage{url}
\usepackage[marginal]{footmisc}
\IEEEoverridecommandlockouts
\usepackage{cite}
\usepackage{amsmath,amssymb,amsfonts}
\usepackage{algorithmic}
\usepackage{graphicx}
\usepackage{subfigure}
\usepackage{textcomp}
\usepackage{xcolor}
\usepackage{balance}
\usepackage{enumerate}
\usepackage{tcolorbox}
\usepackage{colortbl}
\usepackage{tabularx}
\usepackage{longtable}
\usepackage{framed}
\def\BibTeX{{\rm B\kern-.05em{\sc i\kern-.025em b}\kern-.08em
    T\kern-.1667em\lower.7ex\hbox{E}\kern-.125emX}}
\begin{document}

\title{Symptoms of Architecture Erosion in Code Reviews: A Study of Two OpenStack Projects}

\author{
    \IEEEauthorblockN{Ruiyin Li$^{1,2}$, Mohamed Soliman$^2$, Peng Liang$^{1*}$\thanks{\indent This work was funded by the National Key R\&D Program of China with Grant No. 2018YFB1402800 and the NSFC with Grant No. 62172311. The authors acknowledge the support from the China Scholarship Council.}, Paris Avgeriou$^2$}
    \IEEEauthorblockA{$^1$ School of Computer Science, Wuhan University, Wuhan, China}
    \IEEEauthorblockA{$^2$ Department of Mathematics and Computing Science, University of Groningen, Groningen, The Netherlands}
    \IEEEauthorblockA{\{\href{mailto:ryli_cs@whu.edu.cn}{ryli\_cs}, \href{mailto:liangp@whu.edu.cn}{liangp}\}@whu.edu.cn, \{\href{mailto:m.a.m.soliman@rug.nl}{m.a.m.soliman}, \href{mailto:p.avgeriou@rug.nl}{p.avgeriou}\}\href{mailto:@rug.nl}{@rug.nl}}
}


\maketitle

\begin{abstract}
The phenomenon of architecture erosion can negatively impact the maintenance and evolution of software systems, and manifest in a variety of symptoms during software development. While erosion is often considered rather late, its symptoms can act as early warnings to software developers, if detected in time. In addition to static source code analysis, code reviews can be a source of detecting erosion symptoms and subsequently taking action. In this study, we investigate the erosion symptoms discussed in code reviews, as well as their trends, and the actions taken by developers. Specifically, we conducted an empirical study with the two most active Open Source Software (OSS) projects in the OpenStack community (i.e., Nova and Neutron). We manually checked 21,274 code review comments retrieved by keyword search and random selection, and identified 502 code review comments (from 472 discussion threads) that discuss erosion. Our findings show that (1) the proportion of erosion symptoms is rather low, yet notable in code reviews and the most frequently identified erosion symptoms are architectural violation, duplicate functionality, and cyclic dependency; (2) the declining trend of the identified erosion symptoms in the two OSS projects indicates that the architecture tends to stabilize over time; and (3) most code reviews that identify erosion symptoms have a positive impact on removing erosion symptoms, but a few symptoms still remain and are ignored by developers. The results suggest that (1) code review provides a practical way to reduce erosion symptoms; and (2) analyzing the trend of erosion symptoms can help get an insight about the erosion status of software systems, and subsequently avoid the potential risk of architecture erosion.
\end{abstract}

\begin{IEEEkeywords}
Architecture Erosion, Code Review, Open Source Software, OpenStack
\end{IEEEkeywords}

\section{Introduction}\label{S:Introduction}
Along evolution, software systems tend to exhibit a (growing) gap between the intended and the implemented architecture. This phenomenon is commonly referred to as \textit{architecture erosion} \cite{Perry1992ffs} and is often described using diverse terms (e.g., architectural decay and degradation) \cite{li2021uae}. Architecture erosion has a negative impact on the maintenance and evolution of a software system, by causing architectural inconsistencies (e.g., defective interfaces that hinder interaction with outside components) and degradation of software quality (e.g., components that are highly resistant to change) \cite{li2021uae}. However, architecture erosion is typically considered rather late in the process, and especially after it has severely impacted the system \cite{Garcia2021fad}; this makes it hard or sometimes even impossible to handle erosion. Therefore, it is wise to detect and repair architectural erosion as early as possible. 
One way to achieve that is by looking for \textit{symptoms} of erosion in the early phases.

Researchers have identified a number of \textit{architecture erosion symptoms}, which indicate the presence of the aforementioned gap between intended and implemented architecture (see Section \ref{S:Background_a}).
Such symptoms include common violations of design decisions and principles (e.g., violations of inter-module communication rules in a layered architecture \cite{DeSilva2012csa}) and structural issues (e.g., architectural smells \cite{Le2018aes}, \cite{Fontana2016aer}, \cite{Macia2012otr}, rigidity and brittleness of systems \cite{Martin2000dpdp}, \cite{Ayyaz2015aff}). The occurrence of such symptoms can act as early warnings for software engineers to tackle architecture erosion (e.g., by refactoring)  \cite{li2021uae}, \cite{Ali2018acs}. 

Previous studies have investigated approaches, mostly using static source code analysis \cite{Fontana2016aer}, \cite{Le2018aes}, for identifying erosion symptoms \cite{li2021uae}, \cite{Fontana2016aer}, \cite{Le2018aes}, \cite{Macia2012otr}.
However, existing code analysis tools may not be sufficient to accurately identify the wide range of erosion symptoms \cite{Lenhard2017acs}, \cite{Azadi2019asd}. For instance, existing tools (e.g., Sonargraph, Arcan) cannot identify certain types of architectural smells (e.g., ambiguous interface, multipath dependency) \cite{Azadi2019asd}. Moreover, several architectural smells manifest at levels that are out of scope of existing techniques and tools, such as package-level and service-level (including microservices) smells \cite{Mumtaz2021SMS_ASs}. 

While certain erosion symptoms are hard to detect using source code analysis tools, there are other sources that could be tapped for this purpose, such as code reviews.
Code review is a widely used practice to manually find defects in code and improve code quality \cite{Bacchelli2013eoa}. Besides, code reviews can improve the quality attributes of the software \cite{Morales2015dcr} and also result in improvements at the architecture level \cite{Paixao2019tic}. 
Identifying erosion symptoms from code reviews can be complementary to using source code analysis tools; this would enhance the accuracy of detecting erosion symptoms. 
To the best of our knowledge, this data source for detecting erosion symptoms has not been investigated before. In fact, little is known about whether erosion symptoms are widely discussed (if at all) during code reviews and how developers deal with them. 


In this study we aim at empirically investigating architecture erosion symptoms that are discussed during the code review process, as well as their evolution trend, and the actions taken by developers on dealing with the erosion symptoms. 
Identifying erosion symptoms through code review could support detecting architecture erosion early in software systems, that would otherwise not be captured by static analysis of source code. Regarding the scope, we focus on violation and structural symptoms of architecture erosion (see Section \ref{S:Background_a}), as these two types are the most frequently discussed in the literature according to our recent study \cite{li2021SMS} (see Section \ref{S:relatedwork_c}). 

To achieve our goal, we manually identified 502 code review comments (contained in 472 discussion threads) related to the symptoms of architecture erosion. The key contributions of this work are summarized as follows:
\begin{itemize}
    \item We constructed a benchmark dataset containing violation and structural symptoms of architecture erosion for further use by researchers (see the replication package \cite{Replication}).
    \item We manually identified the most frequently discussed architecture erosion symptoms from code reviews and provided a taxonomy of the erosion symptoms.
    \item We further analyzed the trends of the erosion symptoms and the actions taken by developers in dealing with the symptoms.
\end{itemize}

The remainder of this paper is organized as follows. Section \ref{S:Background} describes the background of this work. Section \ref{S:Method} elaborates on the research questions and the study design. Section \ref{S:Results} presents the results, which are subsequently discussed in Section \ref{S:Discussion}. Section \ref{S:Implication} details the implications of the study results for researchers and practitioners and Section \ref{S:Threats} discusses the threats to the validity of the study. Section \ref{S:relatedwork} introduces related work and Section \ref{S:Conclusions} concludes with future directions.

\section{Background}\label{S:Background}

\subsection{Code Review Process}\label{S:Background_b}
Code review is the process of analyzing code submitted by developers to judge whether it is suitable to be integrated into the code base. It is meant as a lightweight practice of code inspection \cite{Fagan1976dci} and is applied in both open source and industrial projects. Code review can be conducted in different ways; for example, OSS communities use not only informal code review processes through online communication (e.g., mailing lists or issue trackers), but also employ formal code review processes supported by tools (e.g., Gerrit\footnote{\url{https://www.gerritcodereview.com/}})~\cite{Beller2014mcr}.

\begin{figure}[htb]
	\centering
	\includegraphics[width=0.48\textwidth]{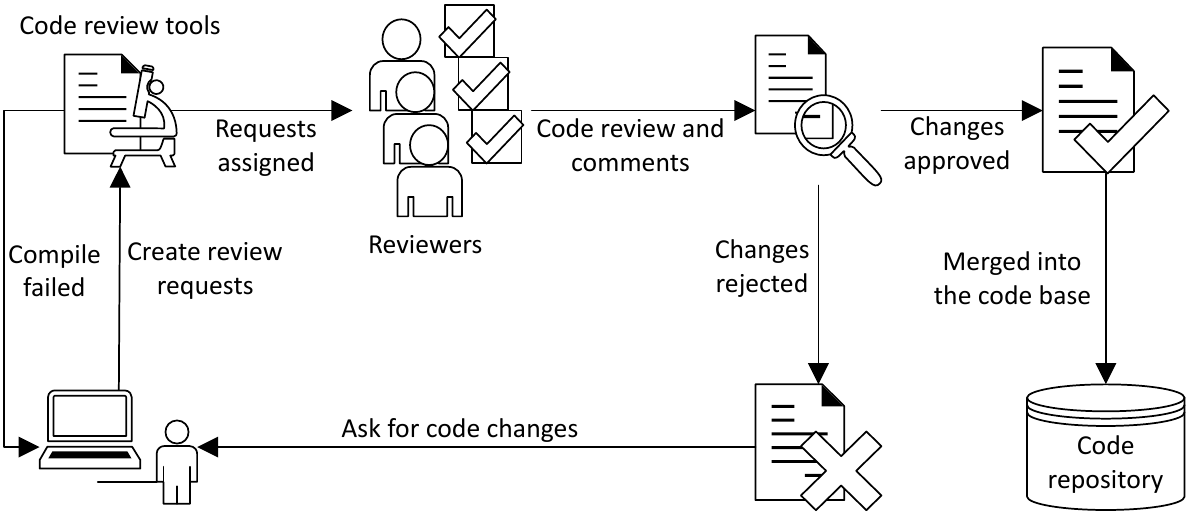}
    	\caption{An overview of the code review process.}\label{F:Overview_of_code_review}
\end{figure}

Code review tools (e.g., Gerrit) are increasingly being adopted in both industrial and OSS projects \cite{Sadowski2018mcr}, \cite{Bosu2015cuc}, \cite{Morales2015dcr}. Figure \ref{F:Overview_of_code_review} shows an overview of the code review process that can be conducted iteratively. A developer creates code review requests and submits the code changes (e.g., patches, bug fixes) to the code review tools where static source code analysis can be automatically conducted for checking errors in code (e.g., compilation errors). After passing the automated analysis, the review requests are assigned to reviewers, which subsequently submit feedback to approve or reject the integration of the code changes into the code base. If the reviewers find defects in the submitted code, they will reject the code changes along with their feedback to the developers. The review process iterates until either the reviewers approve (status ``\textit{MERGED}") or reject (status ``\textit{ABANDONED}") the code changes.

\subsection{Architecture Erosion Symptoms}\label{S:Background_a}

Architecture erosion reflects the deviation of the implemented architecture from the intended architecture over time \cite{Perry1992ffs}, and manifests in a variety of symptoms during software development. A symptom is a partial sign or indicator of the emergence of architecture erosion. Structural and violation symptoms are the most widely discussed erosion symptom types in the literature \cite{li2021SMS} (see Section \ref{S:relatedwork_c}). 
In this work we focus on these two types of symptoms and collectively refer to them as \textbf{architecture erosion symptoms}. 

\textbf{Structural symptoms} denote various structural problems in software architecture. For instance, Le \textit{et al}. \cite{Le2016rad} found that architectural smells can be regarded as structural symptoms affecting the sustainability of software systems. Herold \textit{et al}. \cite{Herold2015dvc} reported that certain structural anti-patterns are symptoms of architecture erosion.

\textbf{Violation symptoms} refer to architectural violations during software development, such as violations of prescribed design decisions (e.g., layered pattern), principles (e.g., encapsulation), or constraints (e.g., uniform interface). For example, prior studies \cite{Fontana2016aer}, \cite{Macia2012otr}, \cite{Mair2014tfa} revealed that architectural violations (e.g., violations of intended design decisions), can be used to indicate the presence of architecture erosion.


\section{Study Design}\label{S:Method}

\subsection{Research Questions}
We aim at investigating the symptoms of architecture erosion that were identified in code reviews, and analyze the distribution and trend of the erosion symptoms, as well as how developers deal with those symptoms. To achieve this goal, we defined three Research Questions (RQs):

\begin{center}{\fbox{\parbox{0.475\textwidth}{\textbf{RQ1: Which symptoms of architecture erosion are frequently identified in code reviews?}}}}\end{center}

This RQ aims at identifying which types of erosion symptoms are frequently discussed during the code review process. Answering this RQ can help to understand erosion symptoms that occur in practice but might be ignored by code analysis tools. The results can also help researchers and practitioners to devise tools and propose approaches for the automated identification of these frequently discussed symptoms.

\begin{center}{\fbox{\parbox{0.475\textwidth}{\textbf{RQ2: How do architecture erosion symptoms identified in code reviews evolve along time?}}}}\end{center}

This RQ aims at investigating how the discussion of architecture erosion symptoms changes along software evolution. That can help to understand to what extent architecture erosion accelerates or slows down. Answering this RQ can provide insights into the evolution of erosion symptoms throughout the code review process, and consequently contribute to understanding the sustainability and stability of architecture.

\begin{center}{\fbox{\parbox{0.475\textwidth}{\textbf{RQ3: Do the architecture erosion symptoms identified in code reviews get fixed in subsequent code changes?}}}}\end{center}

This RQ aims at investigating what actions developers might take after erosion symptoms are identified during the code review process, such as fixing the identified issues or ignoring them. Answering this RQ can help to reveal whether the identified erosion symptoms have an impact on code changes, and shed light on to what degree code reviews can result in removing the identified symptoms of architecture erosion and subsequently improving software quality.

\subsection{Data Collection and Analysis}\label{S:Study Process}
\begin{figure*}[t]
	\centering
	\includegraphics[width=0.95\textwidth]{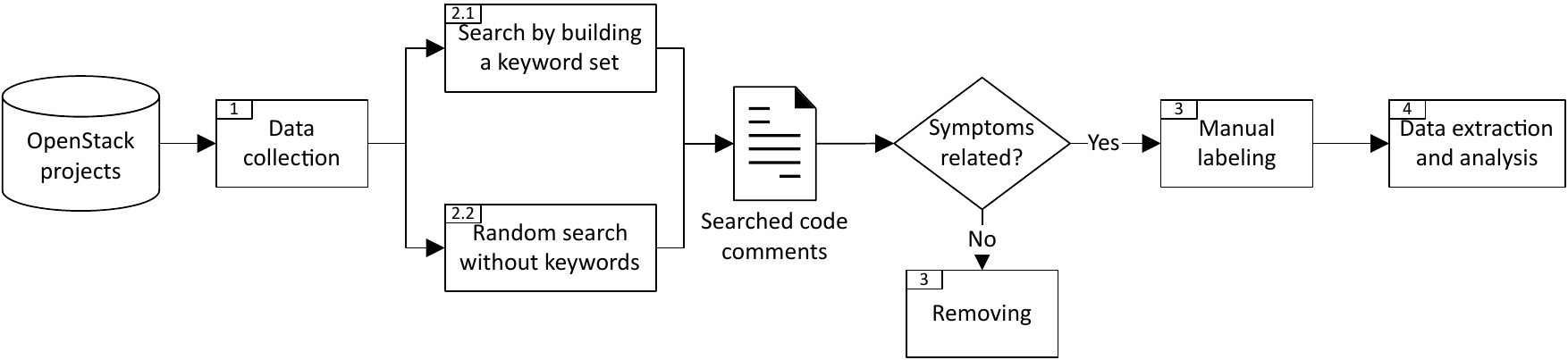}
    	\caption{An overview of the data collection and analysis process.}\label{F:Overview}
\end{figure*}

Figure \ref{F:Overview} shows an overview of the data collection and analysis process. First, we employed a keyword-based approach to mine relevant comments that could potentially discuss erosion symptoms. Of course, the retrieved code review comments that contain the keywords, may not actually discuss erosion symptoms. Thus, we manually checked the retrieved code review comments to identify the discussions that contain erosion symptoms. In addition, we conducted a random selection of the review comments that do not contain any keywords as a supplement to the keyword-based approach. The whole process contains four steps as explained below. All the data and scripts are available at a replication package \cite{Replication}.

\noindent\textbf{Step 1: Data Collection}

The first step is to select software projects and collect code review data from publicly available OSS repositories. OpenStack\footnote{\url{https://www.openstack.org/}} is a widely-used open source cloud software platform, with which many organizations (e.g., IBM) collaboratively develop applications for cloud computing. The code review process in OpenStack is managed by the review tool Gerrit, and the code review data can be accessed through REST API\footnote{\url{https://gerrit-review.googlesource.com/Documentation/rest-api.html}}. We selected two of the projects of the OpenStack platform: Nova (a controller for providing cloud virtual servers) and Neutron (providing networking as a service for interface devices). These are the largest projects within OpenStack, and they have been actively developed over the last seven years with rich code review data. Moreover, the two projects have been used for understanding the detection of code-related quality issues (e.g., code smells \cite{Han2021ucs}). Thus, they might also contain the discussions on architecture erosion symptoms. We employed Python scripts to automatically mine code review data from the two projects between Jan 2014 and Dec 2018 (see Table \ref{T:projects}). We organized the data in a structured way and stored them in MongoDB. 

\begin{table}[htb]
    \footnotesize
    \centering
    \caption{An overview of the subject projects.}
    \label{T:projects}
    \begin{tabular}{|m{8mm}<{\centering}|m{25mm}<{\centering}|m{10mm}<{\centering}|m{10mm}<{\centering}|m{13mm}<{\centering}|}
    \hline
    \textbf{Project} & \textbf{Domain} & \textbf{Language} & \textbf{\#Code Changes} & \textbf{\#Comments}\\\hline
    Nova & Virtual server management & Python & 22,762 & 156,882\\\hline
    Neutron & Network connectivity & Python & 15,256 & 152,429\\\hline
    \end{tabular}
\end{table}


\noindent\textbf{Step 2.1: Retrieve by Building a Keyword Set}

While the selected projects in Table \ref{T:projects} have thousands of comments, a large number of review comments might not be related to architectural issues \cite{Paixao2019tic} and it is inefficient to manually check each review comment. Therefore, we decided to search for the discussions on erosion symptoms through associated keywords. We first needed to determine relevant keywords that commonly occur in the discussions on erosion symptoms. In a recent empirical study on architecture erosion \cite{li2021uae}, Li \textit{et al}. found that developers prefer to use certain common terms (e.g., erosion, decay, degradation) to describe the architecture erosion phenomenon in Q\&A websites. Therefore, we conducted a trial search by using the terms identified in \cite{li2021uae} (e.g., decay, erode, erosion, degrade, degradation, deteriorate, deterioration). However, we found that these description terms of architecture erosion were rarely used in code reviews. A possible explanation is that reviewers, during the code review process: a) focus more on specific code changes rather than general architectural concepts, and b) use terms related to specific architectural issues (e.g., a specific constraint violation).

To effectively locate the specific types of erosion symptoms in code reviews, we needed to establish a keyword set related to violation and structural symptoms of architecture erosion (see Section \ref{S:Background_a}). In Table \ref{T:symptoms}, we selected common violation and structural symptoms according to previous studies \cite{Azadi2019asd}, \cite{Le2016rad}, \cite{Oizumi2019otd}, \cite{Ganesh2013tpc}, and formulated an initial keyword set. For this set, we preferred to be inclusive rather than exclusive and thus included all terms mentioned in the aforementioned studies.

Considering that the effectiveness of keyword-based text mining techniques highly relies on the set of keywords, we worked towards refining this initial set of keywords. Specifically, we followed the iterative approach proposed by Bosu \textit{et al}. \cite{Bosu2014itc}; similarly to our goal, Han \textit{et al}. had established a set of keywords in order to mine code smell discussions in code reviews \cite{Han2021ucs}. In this study, we mined code reviews and change logs of code changes (that contain status information ``\textit{MERGED}" and ``\textit{ABANDONED}") with the purpose of identifying relevant discussions on architecture erosion symptoms. The approach includes the following steps:

\begin{enumerate}
    \item We prepared an initial set of keywords as mentioned above.
    \item We searched using the initial keyword set in the collected review comments and built a corpus by collecting the relevant comments that encompass at least one keyword from the initial set of keywords (e.g., ``violation", ``inconsistent").
    \item We processed the retrieved code review comments which contain at least one keyword in our keyword set and removed English stopwords, punctuation, and numbers. That results in a list of tokens.
    \item We conducted a stemming process (using SnowballStemmer from the NLTK toolkit \cite{Bird2010nlp}) to obtain the stem of each token (e.g., ``architecture" and ``architectural" have the same token ``architectur'').
    \item We built a document-term matrix \cite{tan2016idm} from the corpus, and found the additional words that co-occur frequently with each of our initial keywords (co-occurrence probability of 0.05 in the same document).
    \item We manually checked the list of frequently co-occurring additional words to determine whether the newly discovered words should be added into the keyword set.
\end{enumerate}

After performing the aforementioned steps, we found that no additional keywords co-occurred with the initial keywords based on the co-occurrence probability of 0.05 in the same document; this is similar to previous studies \cite{Han2021ucs} and \cite{Bosu2014itc}. Thus, we believe that the initial keyword set is sufficient to mine the violation and structural symptoms of architecture erosion from code review comments. In total, 20,211 code review comments were retrieved from the two projects. 

\begin{table}[htb]
    \footnotesize
    \centering
    \caption{Symptoms of architecture erosion used in this study}\label{T:symptoms}
    \begin{tabular}{|p{32mm}|p{48mm}|}
    \hline
    \cellcolor{gray}\textbf{Violation symptom} & \cellcolor{gray}\textbf{Keywords} \\\hline
    Architecture violation & architecture, architectural, layer, design, violate, violation, deviate, deviation\\\hline
    Architecture inconsistency & inconsistency, inconsistent, consistent, mismatch, diverge, divergence, divergent\\\hline
    Constraint violation & rule, constraint, violate, violation\\\hline
    \cellcolor{gray}\textbf{Structural symptom} & \cellcolor{gray}\textbf{Keywords}\\\hline
    General terms of architectural smells & architecture, architectural, structure, structural, smell, antipattern, anti pattern, anti-pattern, defect\\\hline
    Cyclic dependency & cycle, cyclic, circular, dependence, dependency\\\hline
    Unnecessary dependency & unnecessary, dependency\\\hline
    Obsolete functionality & obsolete, unused\\\hline
    Ambiguous interface & ambiguous, interface\\\hline
    Unused interface and unused brick & unused, interface, brick\\\hline
    Sloppy delegation & sloppy, delegation\\\hline
    Brick functionality overload & brick, overload\\\hline
    Duplicate functionality & duplicated, clone, copypasted, redundant, copied, overload\\\hline
    Scattered functionality	& scattered\\\hline
    \end{tabular}
\end{table}

\noindent\textbf{Step 2.2: Random Selection of Review Comments without Keywords}

It is possible that reviewers do not use the keywords in Table \ref{T:symptoms} when they described erosion symptoms during the code review process. Thus, we conducted a supplementary search by randomly selecting review comments that do not contain any keywords in Table \ref{T:symptoms} from the rest of the review comments of the two OpenStack projects. We randomly selected 1063 review comments out of the remaining 289,100 review comments (i.e., confidence level of 95\% and margin of error of 3\% \cite{israel1992dss}). Then, we manually checked the code review comments and retrieved 7 review comments related to erosion symptoms.

\noindent\textbf{Step 3: Data Filtering and Labeling}

The search process retrieved a large number of code review comments through keywords, but these still may largely contain irrelevant results. In other words, there are review comments that are not related to discussions on erosion symptoms but contain keywords, such as code snippets, links, and variable names with keywords (e.g., ``\texttt{attr\_interface}"). The irrelevant results were subsequently removed manually. To ensure we have the same understanding of the erosion symptoms when performing the manual removal, we conducted a pilot data filtering and labeling; for that, we used 50 comments randomly-selected from the retrieved comments, which were checked by the first and second authors, independently. To measure the inter-rater agreement between the first two authors, we calculated the Cohen's Kappa coefficient \cite{Cohen1960aca} of the pilot and got an agreement of 0.898. During the formal data filtering and labeling process, the first author conducted the data filtering and labeling, and the second author reviewed and checked the results. Any disagreements on the comments were discussed between the two authors until a consensus was reached. In total, we collected 502 code review comments (from 472 discussion threads) that contain discussions on erosion symptoms from the two OpenStack projects. Note that, a discussion thread can contain more than one code review comments that pertain to the same symptom.

\noindent\textbf{Step 4: Data Extraction and Analysis}

The data items in Table \ref{T:data item} are used to extract relevant data from the two OpenStack projects for answering the RQs. The first author extracted the data which was subsequently reviewed by the second author. To mitigate personal bias, any disagreements were discussed between the first and second authors to reach a consensus. The first author then rechecked the extraction results of all the code reviews to ensure the correctness of the extracted data. Regarding the data analysis, we used descriptive statistics to analyze the extracted comments and refined the existing classifications in the literature.

\begin{table}[htb]
    \footnotesize
    \centering
    \caption{Mapping between the extracted data items and RQs}\label{T:data item}
    \begin{tabular}{|c|p{22mm}|p{40mm}|p{4mm}<{\centering}|}\hline
    \textbf{\#} & \textbf{Data item} & \textbf{Description} & \textbf{RQ}\\\hline
     D1 & Comment & The comments from reviewers on source code & RQ1\\\hline
     D2 & Code review URL & The URL link of the code review comments and changes & RQ1\\\hline
     D3 & Change ID & The unique id of each code change & RQ1\\\hline
     D4 & Comment timestamp & The timestamp of the code review comments & RQ2\\\hline
     D5 & Code change status & The status of each code change in the change logs & RQ3\\\hline
    \end{tabular}
\end{table}

\section{Results}\label{S:Results}
\subsection{Results of RQ1}\label{S:Results_RQ1}

\begin{figure}[t]
	\centering
	\includegraphics[width=0.5\textwidth]{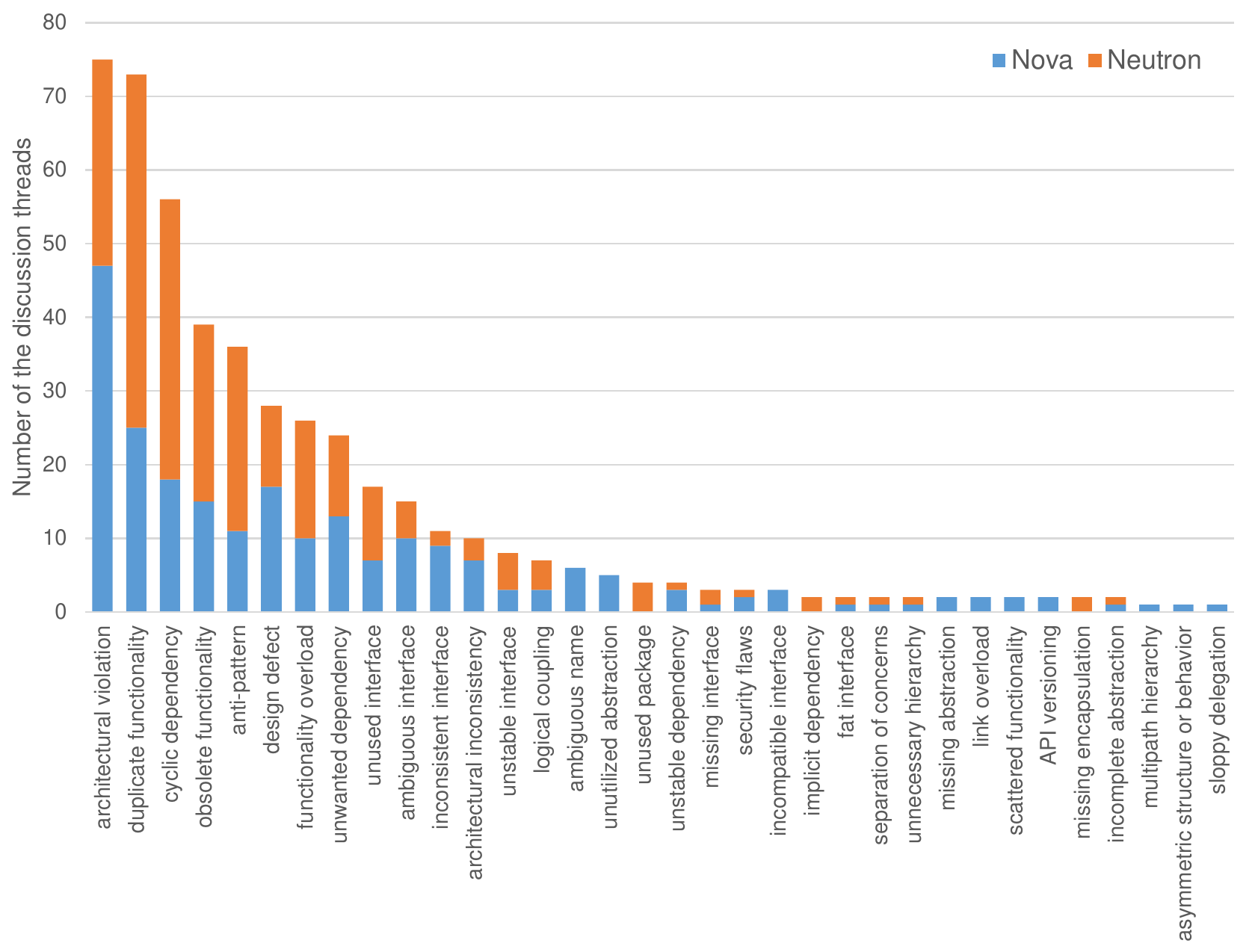}
    	\caption{Number of the discussion threads on architecture erosion symptoms in Nova and Neutron.}\label{F:symptoms}
\end{figure}

In total, we identified 502 code review comments (contained in 472 discussion threads) related to architecture erosion symptoms. As mentioned in Section \ref{S:Study Process}, each discussion thread reflects one symptom but may contain more than one comment. Figure \ref{F:symptoms} shows the distribution of the identified erosion symptoms in the code reviews from Nova and Neutron. \textit{Architectural violation} is the most frequently identified symptom with 75 (15.9\%) discussion threads; this symptom entails that the reviewers perceived the implemented architecture to be violating the intended architecture during the development process. \textit{Duplicate functionality} \cite{Le2016rad} is a common architectural smell that comes second with 71 (15.5\%) discussion threads, followed by \textit{cyclic dependency} with 56 (11.9\%) discussion threads. In addition, 39 (8.3\%) discussion threads belong to \textit{obsolete functionality}.

Considering the two categories of symptoms, i.e., violation symptoms and structural symptoms, the former amounts to 18.0\% and comprises two categories: \textit{architectural violation} (75 discussion threads) and \textit{architectural inconsistency} (10 discussion threads). \textit{Architectural violation} entails the violation of architectural constraints (e.g., layered pattern) or design rules (e.g., encapsulation). For example, one reviewer mentioned that ``\textit{this feels like a layer violation to be re-using the privsep stuff from os-brick here}"\footnote{\url{https://review.opendev.org/c/openstack/nova/+/312488}}. \textit{Architectural inconsistencies} reflect a mismatch between the implemented and the intended architecture (e.g., inconsistent dependency). For example, one developer responded: ``\textit{ok I will remove that, there is inconsistent between agent implementations}"\footnote{\url{https://review.opendev.org/c/openstack/neutron/+/195439}}.

Compared to violation symptoms, structural symptoms (82.0\%) are much more frequently discussed in code reviews. In the following, we refined the existing classification (e.g., \cite {Azadi2019asd}, \cite{Le2016rad}) and provide a taxonomy of the most frequently discussed structural symptoms. Due to space limitations, we list the major structural symptoms in each category, while the rest can be found in the replication package \cite{Replication}.
\begin{enumerate}[(1)]
    \item \textbf{Functionality-related smells}: Four types of structural symptoms contained in 140 discussion threads are related to functionality, namely \textit{duplicate functionality}, \textit{obsolete functionality}, \textit{functionality overload}, and \textit{scattered functionality}.
    \begin{itemize}
        \item \textit{Duplicate functionality} (73 discussion threads) concerns the replication of the same functionality among components. Ignoring the other components when altering the functionality of one component might cause architectural problems. For example, one reviewer commented: ``\textit{this is duplicated in the other module - we could pull this into the common module as a helper method}"\footnote{\url{https://review.opendev.org/c/openstack/nova/+/229964}}.
        \item \textit{Obsolete functionality} (39 discussion threads) refers to unused and invalid functionality that can increase system complexity. For example, one reviewer stated ``\textit{oh, this module is not necessary at all now because we have already removed legacy v2 API code}"\footnote{\url{https://review.opendev.org/c/openstack/nova/+/292473}}.
        \item \textit{Functionality overload} (26 discussion threads) occurs when a component undertakes an excessive amount of functionality. For example, as described by one reviewer ``\textit{since this module is overloaded by tons of code, it makes sense for me to get it rid of all responsibilities that we can do}"\footnote{\url{https://review.opendev.org/c/openstack/nova/+/282580}}.
    \end{itemize}
    
    \item \textbf{Dependency-related smells}: We identified five types of structural symptoms contained in 91 discussion threads related to dependency issues, namely \textit{cyclic dependency}, \textit{unwanted dependency}, \textit{unstable dependency}, \textit{implicit dependency}, \textit{link overload}, \textit{unnecessary hierarchy}, and \textit{multipath hierarchy}.
    \begin{itemize}
        \item \textit{Cyclic dependency} (56 discussion threads) occurs when two or more components directly or indirectly interact with each other to form a circular chain. For example, as one reviewer pointed out ``\textit{the removal here is a good example how the restructuring resolves the cyclic dependency issue}"\footnote{\url{https://review.opendev.org/c/openstack/nova/+/250907}}.
        \item \textit{Unwanted dependency} (24 discussion threads) denotes the dependencies that are obsolete, duplicate, or unnecessary between components. As one reviewer stated ``\textit{I think it's a wrong dependency, [...]}"\footnote{\url{https://review.opendev.org/c/openstack/neutron/+/87841}}.
    \end{itemize}
    
    \item \textbf{Interface-related smells}: We identified nine types of structural symptoms contained in 69 discussion threads related to interfaces, namely \textit{unused interface}, \textit{ambiguous interface}, \textit{inconsistent interface}, \textit{unstable interface}, \textit{logical coupling}, \textit{missing interface}, \textit{incompatible interface}, \textit{fat interface}, \textit{API versioning}, and \textit{sloppy delegation}.
    \begin{itemize}
        \item \textit{Unused interface} (17 discussion threads) occurs when an interface is not utilized. For example, one reviewer suggested ``\textit{this is a weird interface. MonitorBase doesn't seem to be used very much yet, is this a good opportunity to remove this abstract method and replace it with get\_metrics?}"\footnote{\url{https://review.opendev.org/c/openstack/nova/+/219153}}.
        \item \textit{Ambiguous interface} (15 discussion threads) appears when an interface has an unclear definition or provides a single or general entry-point (e.g., single parameter) to other components or connectors. For example, one reviewer stated ``\textit{I won't hold up on it, but this is a bit of a confusing interface to me, so I think documenting it thoroughly is important}"\footnote{\url{https://review.opendev.org/c/openstack/nova/+/427902}}.
    \end{itemize}
    
    \item \textbf{General symptoms} refer to general descriptions of the structural symptoms, including \textit{anti-patterns} (36 discussion threads) and \textit{design defects} (28 discussion threads). For example, one reviewer stated ``\textit{This is an antipattern in concurrent programming and we should not do this.}"\footnote{\url{https://review.opendev.org/c/openstack/nova/+/77995}}.
\end{enumerate}

\begin{center}
\fbox{\parbox{0.475\textwidth}{\textbf{Finding 1}: The proportion of architectural erosion symptoms is rather low in code reviews. The most frequently identified symptoms of architectural erosion are \textit{architectural violation}, \textit{duplicate functionality}, and \textit{cyclic dependency}.}}
\end{center}

\subsection{Results of RQ2}\label{S:Results_RQ2}
To answer RQ2 and gain a first insight into the architectural erosion trends of the two OpenStack projects (i.e., Nova and Neutron), we plotted line charts of the discussion threads on erosion symptoms in the two projects between 2014 and 2018.

\begin{figure}[t]
	\centering
	\begin{center}
		\centering
		\includegraphics[width= 0.5\textwidth]{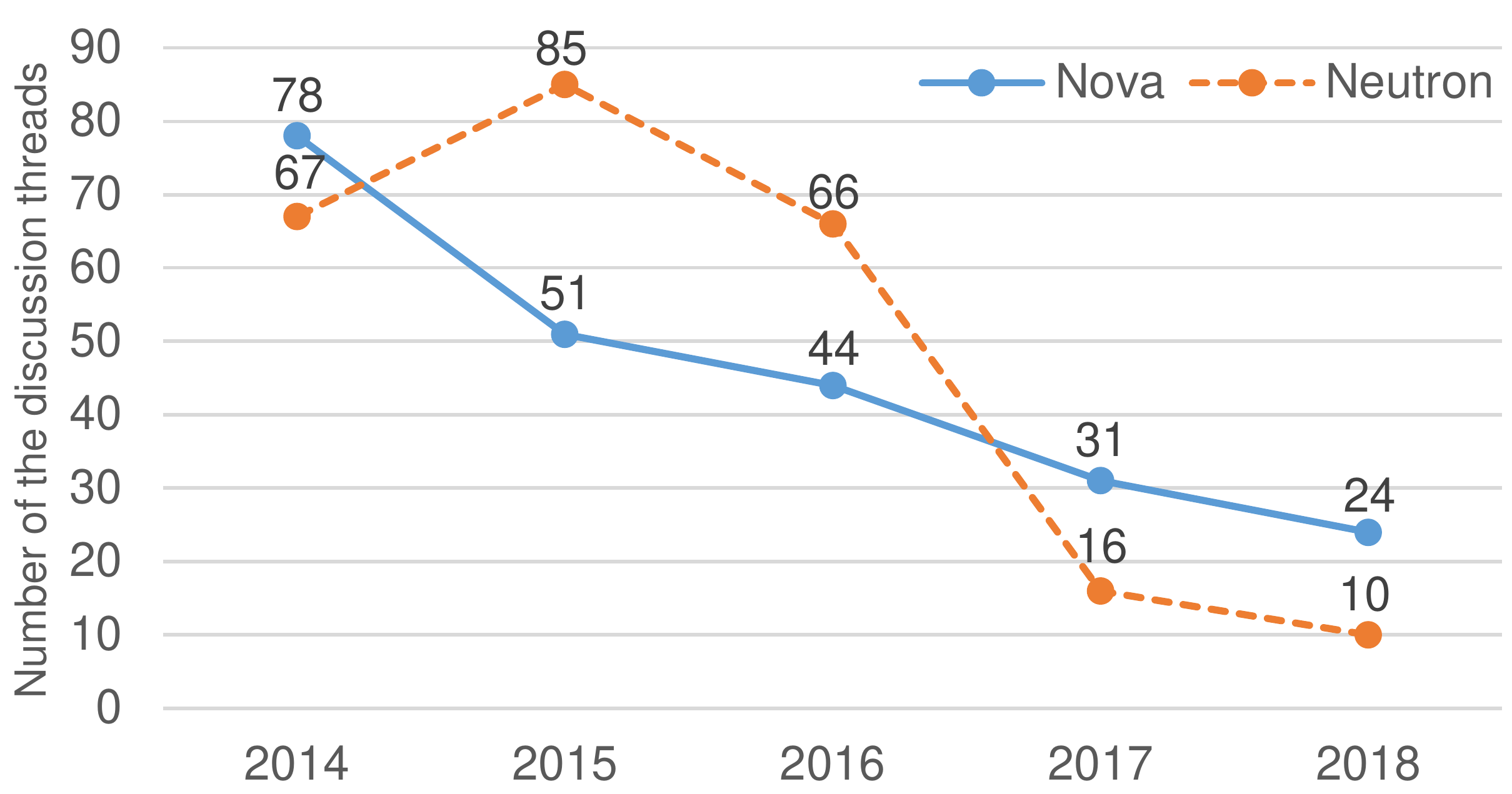}
		\caption{Trends of the discussion threads on architecture erosion symptoms in Nova and Neutron.}\label{F:Year_Trend}
	\end{center}
	\begin{center}
		\centering
        \includegraphics[width= 0.5\textwidth]{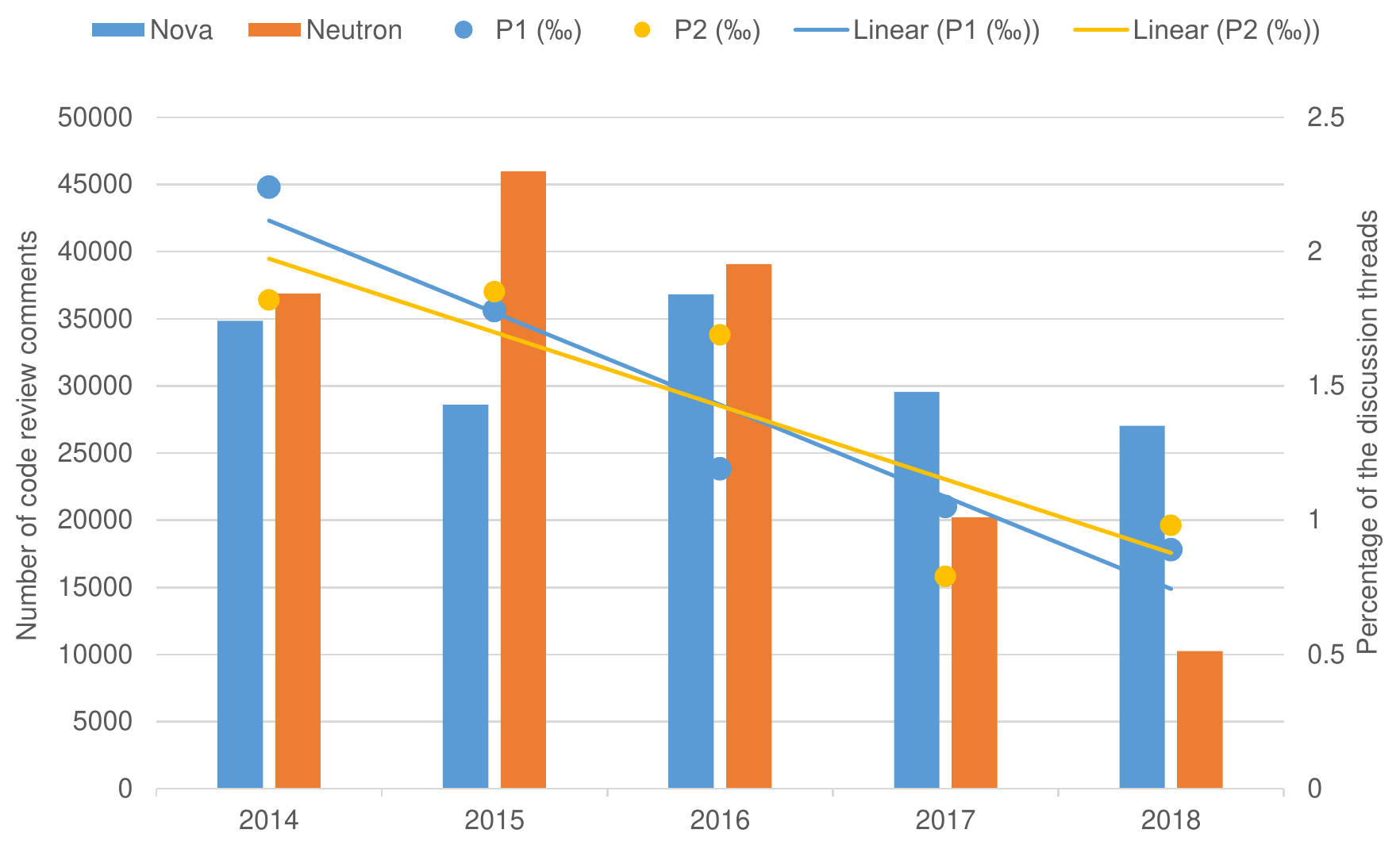}
    	\caption{Percentages of the discussion threads on architecture erosion symptoms in Nova (i.e., P1) and Neutron (i.e., P2).}\label{F:Year_Trend_Percentage}
	\end{center}
\end{figure}

As we can see from Figure \ref{F:Year_Trend}, the numbers of discussion threads on erosion symptoms show a decreasing trend for both Nova and Neutron, as the two projects evolve. For Nova, we can see that the number of erosion symptoms shows a steady downward trend from 2014 to 2018. As for Neutron, the number of erosion symptoms fluctuates and reaches a peak in 2015, and then it follows a decreasing trend in the next three years. One possible reason for the drop in the numbers of erosion symptoms in both projects could be a potential decrease in the numbers of review comments. To check this, we looked at the percentage of erosion symptoms in review comments (see Figure \ref{F:Year_Trend_Percentage}). Although the numbers of review comments fluctuate over time, we found that the percentages of erosion symptoms have a declining tendency through a linear regression analysis. 
Thus, it is reasonable to argue that the architecture of the two OpenStack projects becomes stable over time, as they exhibit fewer erosion symptoms.

\begin{figure*}[t]
	\centering
	\includegraphics[width=0.975\textwidth]{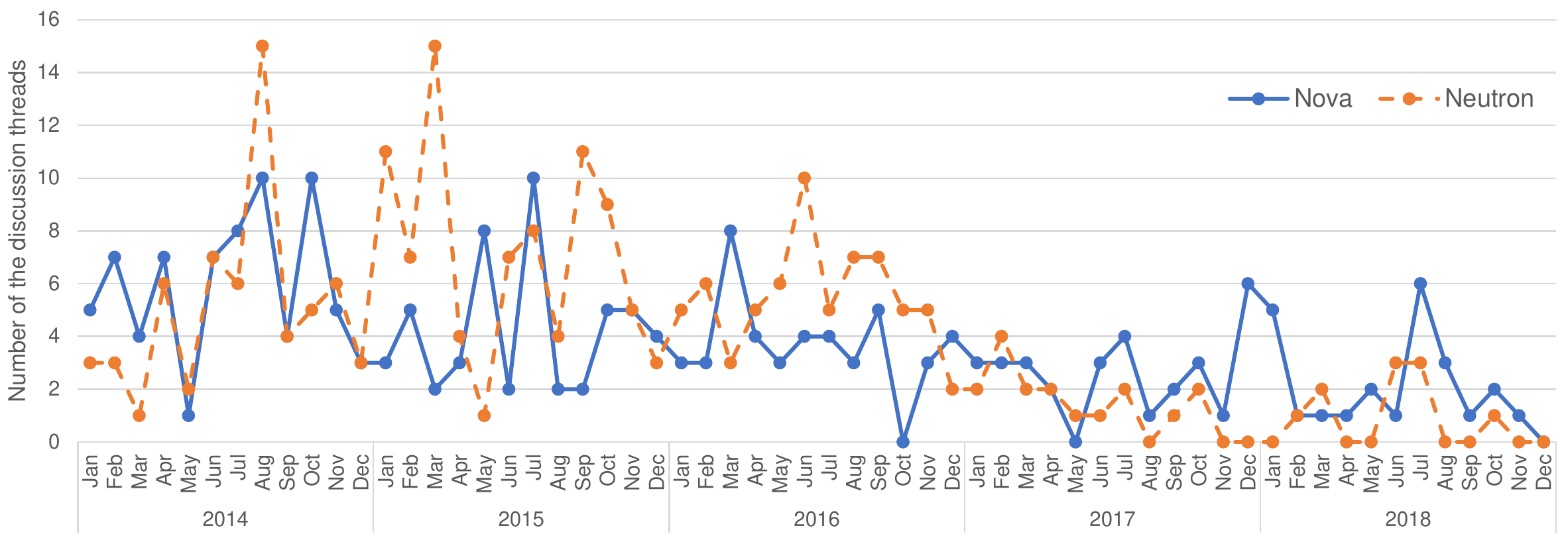}
    	\caption{Monthly distribution of the discussion threads on architecture erosion symptoms per month in Nova and Neutron.}\label{F:Month_Trend}
\end{figure*}

To perform a more detailed analysis, we analyzed the distribution of the numbers of discussion threads on erosion symptoms in the two projects over a period of 60 months between 2014 and 2018 (see Figure \ref{F:Month_Trend}). We can see a fluctuating but downward trend for both Nova and Neutron. 
During the manual check process, we found several reviews that discussed issues concerning both Nova and Neutron, such as connections through interfaces. For example, one reviewer mentioned: ``\textit{I'm still concerned that Nova is not interacting with Neutron through a well defined interface here}"\footnote{\url{https://review.opendev.org/c/openstack/nova/+/275073}}. Interestingly, the two projects have largely similar fluctuating trends of the number of erosion symptoms (see Figure \ref{F:Month_Trend}) along time, since both projects belong to the OpenStack platform, and their erosion symptoms to some extent follow a similar trend.

To statistically verify this trend, we conducted a correlation analysis, using as null hypothesis ($H_0$) that \textit{there is no linear relationship between the number of architecture erosion symptoms in Nova and Neutron}. We chose a p-value 0.05 as a statistical significance threshold, which is commonly used in empirical studies \cite{Shull2007gae}. We used Spearman’s rank correlation \cite{zar1972Spearman} to assess the correlation between the monthly distribution of erosion symptoms in Nova and Neutron. The reason is that Spearman's rank correlation does not have any requirements on the normality of data distribution \cite{pagano2012usbs}. We adopted the classification of correlation coefficient by Marcus and Poshyvanyk \cite{marcus2005ccc}, where a value belonging to [0.3-0.5] denotes a moderate correlation. We calculated a Spearman's rank correlation value of 0.412 (p-value is 0.0007), which indicates a statistically significant positive correlation (p-value $\ll$ 0.01); the correlation of 0.412 is moderate strong. Therefore, we rejected the null hypothesis ($H_0$) and accepted the alternative hypothesis ($H_1$), namely, \textit{there is a linear relationship between the number of architecture erosion symptoms in Nova and Neutron}. 

\begin{center}
\fbox{\parbox{0.475\textwidth}{\textbf{Finding 2}: Both the numbers and percentages of the architecture erosion symptoms in Nova and Neutron show declining tendencies, which suggests that the architecture of the two projects tends to become stable over time. To some extent, the architecture erosion symptoms in Nova and Neutron projects are correlated to each other.}}
\end{center}

\subsection{Results of RQ3}\label{S:Results_RQ3}

\begin{figure*}[t]
	\centering
	\includegraphics[width=0.975\textwidth]{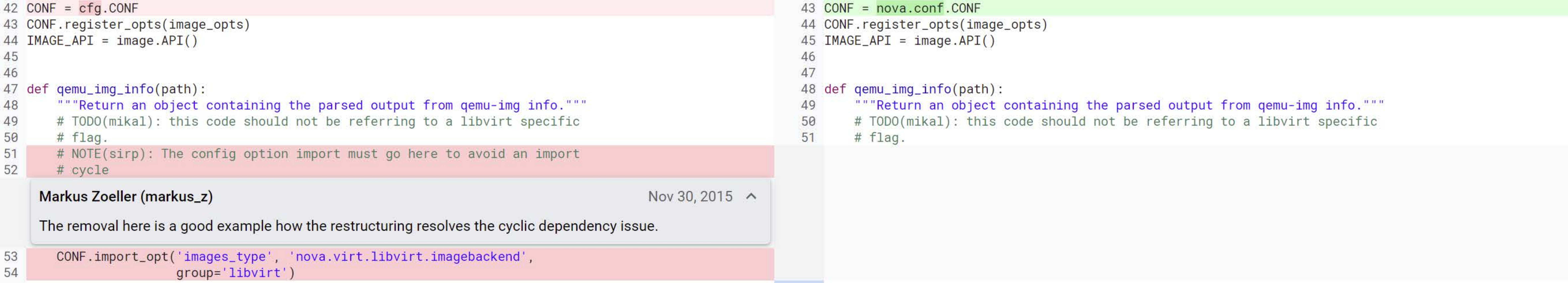}
    	\caption{An example of cyclic dependency remove operation.}\label{F:example}
\end{figure*}

During code review, the review comments can provide suggestions on how to remove architecture erosion symptoms. For example, Figure \ref{F:example} shows a code snippet with a cyclic dependency before the review (left side) and after the dependency was removed (right side). Although the final versions of the submitted code changes can be detected by the status of code changes (status ``\textit{MERGED}" or ``\textit{ABANDONED}"), whether the erosion symptoms get fixed through code changes is not clear. Thus, to gain more insights from the code changes along with their discussions and status, we further investigated and presented the status of code changes in Table \ref{T:Actions}, including 361 ``\textit{MERGED}" and 111 ``\textit{ABANDONED}" code changes.

\begin{table}[htb]
    \footnotesize
    \centering
    \caption{Status of the code changes}\label{T:Actions}
    \begin{tabular}{|l|c|c|}\hline
    \textbf{Status} & \textbf{Count} & \textbf{Sum}\\\hline
    Merged (fixed) & 309 & \multirow{3}{*}{361}\\\cline{1-2}
    Merged (no response) & 49 & ~\\\cline{1-2}
    Merged (no response but fixed) & 3 & ~\\\hline
    Abandoned (verification failed) & 79 & \multirow{3}{*}{111} \\\cline{1-2}
    Abandoned (verification passed but voted to reject) & 30 & ~\\\cline{1-2}
    Abandoned (verification passed but no vote) & 2 & ~\\\hline
    \end{tabular}
\end{table}

``\textit{MERGED}" status denotes that code changes pass the verification (i.e., automated testing) and receive the approval from the reviewers to be integrated into the main repository. The results show that most of the ``\textit{MERGED}" code changes (85.6\%, 309 out of 361) reached an agreement (fixed) after discussions between the reviewers and developers. For example, a reviewer mentioned ``\textit{This is a layering violation. This should be: \texttt{old\_flavor = instance.get\_flavor('old')}}", which was responded by a developer ``\textit{Done}"\footnote{\url{https://review.opendev.org/c/121409}}. 
This example shows that a reviewer pointed out a violation symptom (i.e., layering violation) that needed to be fixed. In addition, we noted that 13.6\% (49 out of 361) of the ``\textit{MERGED}" code changes did not get any response from the developers or reviewers. In other words, while the identified erosion symptoms can be regarded as potential problems, they still remain in the code and are ignored by both developers and reviewers. 

Regarding the code changes marked as ``\textit{ABANDONED}", they either failed their verification, or were voted to be rejected by the reviewers, or both. The results show that most of the ``\textit{ABANDONED}" code changes (71.2\%, 79 out of 111) failed their verification, while 27.0\% of the code changes were rejected by the reviewers after passing the automated testing. Such rejections are mostly due to code changes having bad quality or logical errors; for instance, one reviewer mentioned an issue ``\textit{no, it is not cleaner, it is a design mistake [...]}", and the developer agreed to resubmit a code change ``\textit{ok, I agree.  I will make an update}"\footnote{\url{https://review.opendev.org/c/openstack/nova/+/120309}}. Another common reason for rejecting code changes that passed their verification, is a difference of opinion between developers and reviewers, which leads to reaching an agreement (upon discussion) to re-submit a code change. Moreover, we found that nearly half (45.0\%, 50 out of 111) of the ``\textit{ABANDONED}" code changes did not receive any response from the developers and no actions were taken to cope with the identified erosion symptoms.  

In summary, the results show that code reviews can have a significant and positive impact on removing erosion symptoms: they were either \textit{fixed and merged} or \textit{abandoned} (thus also removing the erosion symptoms). Most (89.6\%, 423 out of 472) of the erosion symptoms were addressed and only 10.4\% (49 out of 472) of the discussion threads that contain erosion symptoms were ignored by the developers and reviewers.

\begin{center}
\fbox{\parbox{0.475\textwidth}{\textbf{Finding 3}: Most code reviews that identify erosion symptoms have a positive impact on removing the architecture erosion symptoms, but a few erosion symptoms remain and are ignored by the developers and reviewers (i.e., they persisted in the ``\textit{MERGED}" code changes). The main reason behind ``\textit{ABANDONED}" code changes is failed verification (i.e., failed automated test).}}
\end{center}

\section{Discussion}\label{S:Discussion}
\subsection{RQ1: Frequently Identified Erosion Symptoms}\label{S:RQ1_Dis}
Our results indicate that the proportion of the review discussions on architecture erosion symptoms is rather low in code reviews (i.e., 2.4\%, 502 out of 21,274 review comments). But they do exist and can be very useful to mine, as they provide an indication of the system's architectural erosion overall. Compared to previous studies, particularly \cite{Azadi2019asd}, \cite{Le2016rad} that only focus on architectural smells and their classification, we also focused on another type of erosion symptoms, that is, architectural violations. Besides, we refined the classification and provided a taxonomy that is comparatively more comprehensive and covers two common types of erosion symptoms (see Section \ref{S:Results_RQ1}). According to the results of RQ1, \textit{architectural violation}, \textit{duplicate functionality}, and \textit{cyclic dependency} are the most frequently identified erosion symptoms. Regarding the structural symptoms of architecture erosion, functionality-related smells (e.g., duplicate functionality) attract more attention from code reviewers, followed by dependency-related and interface-related smells.

We observed that a large number of code review comments concern code-level problems instead of architecture-level problems. In other words, compared to the discussions on code-level issues (e.g., code smells \cite{Han2021ucs}), the percentage of the discussions on architecture erosion symptoms is lower. One obvious reason is that code reviewers usually focus on the submitted code snippets and seldom discuss architectural problems, since they may not really be familiar with the system architecture or they may lack architectural awareness \cite{Paixao2019tic}. Interestingly, many of the discussion threads that contain erosion symptoms are described as bug fixing during the code review process; this may be because code reviewers are not aware that the ``bugs" may undermine the system architecture and make the architectural deviate from the intended architecture. This conjecture corroborates recent findings by Paixao \textit {et al}. \cite{Paixao2019tic} that the majority of ``fixed bugs" affecting the system's behaviour is more than simple bugs throwing an exception. In other words, the code reviews related to architectural changes may be described as code-related changes (e.g., bugs) due to the lack of architectural awareness.

\subsection{RQ2: Trend of Identified Erosion Symptoms}\label{S:RQ2_Dis}
The number of architecture erosion symptoms in the two OpenStack projects (i.e., Nova and Neutron) shows a declining tendency. This indicates that fewer erosion symptoms were discussed during the code review process; as the two projects get mature and stable over time, so does their architecture. This is similar to the findings of Bi \textit{et al}. \cite{Bi2021aic}, who established that architectural discussions and changes decrease after several stable releases. One possible reason is that the architecture of the two projects tends to become stable after several major versions were released. As mentioned in Lehman's Law of software evolution, ``\textit{the quality of a system will appear to be declining during its evolution, unless proactive measures are taken}" \cite{Lehman2002sep}. The declining trend of the erosion symptoms in the two projects implies that positive architectural evolution might take place to improve the structural quality \cite{Uchoa2020hdm} and make the architecture more stable over time.

We also observed that the number of the review comments on erosion symptoms in the two OpenStack projects shows a similar changing trend and are correlated to each other. The two projects (i.e., Nova and Neutron) are important components of the OpenStack cloud computing platform; this entails that there may be many interactions (e.g., couplings of structure and data) between the projects in the same community (e.g., OpenStack in this study), as reflected in the (almost) common evolution of their erosion symptoms. 
Moreover, the finding indicates that when developers change certain components, they should be aware of the potential couplings among the projects in the same platform.

\subsection{RQ3: Impact of Identified Erosion Symptoms}\label{S:RQ3_Dis}
The results of RQ3 show that most of the identified erosion symptoms (89.6\%) were addressed through either being \textit{fixed and merged} or \textit{abandoned} after review votes. This indicates that code reviewers can effectively find the possible architectural issues in code and provide feasible refactoring recommendations to help developers to repair or remove the code snippets with erosion symptoms.
It also indicates that the code review process has a positive impact on architectural improvements and sustainability by removing erosion symptoms in the two OpenStack projects.

However, a few of the identified erosion symptoms (10.4\%) still remained and were ignored by developers and reviewers. One potential reason is that either the identified erosion symptoms did not attract the attention from the developers or different opinions existed between the developers and reviewers about the severity of the symptoms. The remaining erosion symptoms may increase the risk of architecture erosion, ultimately hindering the maintenance and evolution activities in the future. 

\section{Implications}\label{S:Implication}
\subsection{Implications for Researchers}\label{S:Implication_R}
\textbf{Establishing classification models to automatically identify architecture erosion symptoms}. Whether architecture erosion symptoms are accurately identified plays a significant role in preventing architecture erosion and further extending the longevity of systems and their architecture. In this work, the identification process was manual. Researchers can focus on how to precisely and \textit{automatically} identify erosion symptoms from various software artifacts. Furthermore, researchers can attempt to construct prediction models for recommending to \textit{remove} erosion symptoms \cite{Garcia2021fad}. Erosion symptoms can be an early warning of architecture erosion, and the recommended repair activities (e.g., refactorings) can help to reduce the risk of architecture erosion. In this sense, establishing shared and labeled datasets of erosion symptoms based on diverse artifacts (e.g., source code, design documents) is the prerequisite for building models (e.g., classifiers based on deep learning techniques), which can be applied to form a more reliable basis for the identification of erosion symptoms. 

\textbf{Exploring the evolution of erosion symptoms}. The aforementioned findings suggest that researchers can explore the evolution of architecture erosion symptoms for the purpose of preventing architecture erosion. For instance, researchers can measure the density of erosion symptoms in different modules over time and intervene if certain thresholds are reached (e.g., by refactoring). In addition, researchers can investigate the erosion symptoms that are not addressed in order to analyze whether and how the ignored erosion symptoms can have a negative impact on the system over time, as well as the root reasons why developers and reviewers ignored the symptoms.

\subsection{Implications for Practitioners}\label{S:Implication_P}
\textbf{Providing support for managing erosion symptoms}. The findings of this work can provide guidance for developers in conducting refactoring and maintenance activities to deal with architecture erosion. For example, practitioners can pay more attention to those erosion symptoms that have the highest frequency (e.g., duplicate functionality). Moreover, to remove existing erosion symptoms and avoid introducing new erosion symptoms, practitioners can use the information on identified erosion symptoms in different textual artifacts (e.g., code reviews, commits, issues); raising their erosion awareness can help them be more cautious when performing code changes (e.g., adding new features). 

\textbf{Keeping an eye on the remaining and ignored erosion symptoms}. At present, the reasons for developers to accept a solution that will damage the architecture (e.g., sub-optimal design decisions leading to erosion symptoms) or refuse to address architecture erosion symptoms are still a research area under investigation \cite{Paixao2019tic}. To avoid the proliferation of erosion symptoms and mitigate the risk of architecture erosion, we encourage practitioners to systematically assess the severity of the ignored erosion symptoms, and continuously measure the impact of the remaining erosion symptoms (e.g., measure architecture smell size and density \cite{Sharma2020eir}).

\section{Threats to Validity}\label{S:Threats}
We discuss the threats to the validity by following the guidelines proposed by Wohlin \textit{et al}. \cite{Wohlin2012ese}. Internal validity is not discussed, since we did not study causality.

\textbf{Construct Validity} pertains to the connection between the research questions and the objects of our study. One potential threat concerns the selection of the keyword set, namely, whether the keyword set was incomplete. To reduce the risk of missing keywords, we summarized the frequently-used keywords reported in previous studies, and identified the keywords by following the approach proposed by Bosu \textit{et al}. \cite{Bosu2014itc}. Additionally, to further mitigate this threat, we conducted a random selection of review comments that did not contain any keywords.



\textbf{External validity} concerns the generalizability of the study findings. Our study used two OSS projects (i.e., Nova and Neutron) from the same community (i.e., OpenStack) and we have limited the code review data obtained from Gerrit between 2014 and 2018. These factors might restrict the generalizability of our findings in other settings, such as industrial systems, other OSS systems and different time periods. However, considering the popularity and size of the selected projects, we believe that our findings can provide researchers and practitioners an understanding of the common types and trends of architecture erosion symptoms identified in code reviews of large OSS systems, as well as the actions against the erosion symptoms.

\textbf{Reliability} refers to whether the study would yield the same results when other researchers conducted it. To reduce this threat, we ran a pilot data filtering and labeling to eliminate the misinterpretation of the results. The formal data filtering and labeling was conducted by the first author, and reviewed and checked by the second author; we obtained a Cohen's Kappa value \cite{Cohen1960aca} of 0.898. During data extraction and analysis, the data was extracted by the first author and reviewed by the second author. Any disagreements were discussed and addressed during the labeling and manual analysis of code review data. Besides, we specified the process of our study in Section \ref{S:Method} and provided a replication package online \cite{Replication}, which partially mitigate threats to the reliability of the study. 


\section{Related Work}\label{S:relatedwork}
\subsection{Code Review}\label{S:relatedwork_a}
Code review is performed in a variety of ways for different purposes, for example, lightweight tool-based reviews \cite{Bacchelli2013eoa}, checklist-based reviews \cite{Goncalves2020der}, and search-based reviews \cite{Ouni2016spr}, \cite{Bosu2014itc}, \cite{Han2021ucs}. Bacchelli and Bird \cite{Bacchelli2013eoa} investigated the tool-based code review process across different teams at Microsoft. They found that code review can facilitate knowledge transfer among team members and increase the team awareness, while available tools for code review do not always meet developers’ expectations. Han \textit{et al}. \cite{Han2021ucs} conducted an empirical study of code smell detection via code review, using a keyword-based approach to mine code review discussions of two OpenStack projects. They investigated the frequently identified code smells and the corresponding actions taken by developers. Ouni \textit{et al}. \cite{Ouni2016spr} proposed the RevRec approach that formulates the peer code reviewers recommendation problem as a combinatorial search-based optimization problem, which provides decision-making support for code change submitters and reviewers to identify the most appropriate reviewers for code changes. Considering the benefits of code review on software development, we chose code review comments as our data source to empirically investigate the discussions on architecture erosion symptoms among developers during the code review process.

\subsection{Identification of Architecture Erosion Symptoms}\label{S:relatedwork_c}
Macia \textit{et al}. \cite{Macia2012otr} explored the impact of code anomalies on architecture erosion, and their results revealed that certain kinds of early architectural smells (e.g., ambiguous interface, module concern overload) can be regarded as key architecture erosion symptoms and accelerate the erosion of architecture. Uch{\^{o}}a \textit{et al}. \cite{Uchoa2020hdm} analyzed the impact of code review on design degradation evolution. They analyzed various erosion symptoms to investigate the relationships between the density and diversity of symptoms and design degradation. Oizumi \textit{et al}. \cite{Oizumi2019otd} conducted an exploratory study to investigate whether symptoms of structural degradation (e.g., broken modularization, cyclic hierarchy, unutilized abstraction) with higher density and diversity in classes can be used as indicators of the need for root canal refactorings (a refactoring tactic). Their results indicated that certain symptoms might be indeed strong indicators of structural degradation, despite not being removed by refactoring. Mair \textit{et al}. \cite{Mair2014tfa} considered architecture violations as a type of symptom of architecture erosion and investigated how software engineers repaired eroded software systems. Le \textit{et al}. \cite{Le2018aes}, \cite{Le2016rad} mentioned that architectural smells can be regarded as symptoms of architecture erosion and they proposed algorithms and metrics to detect instances of architecture erosion by analyzing the detected smells. Compared to the aforementioned studies that focus on the erosion symptoms by source code analysis (e.g., density of code smells \cite{Uchoa2020hdm}, architectural smells \cite{Le2018aes}, \cite{Le2016rad}), our work investigated the \textit{discussions} on architecture erosion symptoms in code reviews, including the frequently discussed erosion symptoms and their trends, as well as the actions (i.e., \textit{code changes}) taken by developers.

\section{Conclusions and Future Work}\label{S:Conclusions}
To some extent, the trend of architecture erosion symptoms can reflect the trend of system sustainability and stability during evolution \cite{Le2016rad}. 
In this work, to study architecture erosion symptoms in code reviews, we performed an empirical study using the discussions of architecture erosion symptoms in code reviews by collecting and analyzing review comments from the two largest OpenStack projects (i.e., Nova and Neutron). Our findings show that \textit{architectural violation}, \textit{duplicate functionality}, and \textit{cyclic dependency} are the most frequently identified erosion symptoms. The numbers and percentages of review comments on identified erosion symptoms manifest a declining trend in the two OpenStack projects, which indicates that their architecture becomes stable over time. Most of the identified erosion symptoms (89.6\%) were addressed through either being \textit{fixed and merged} or \textit{abandoned} after review votes. This implies that, to some extent, code reviews might have a positive impact on removing erosion symptoms and extending the longevity of systems and their architecture.

Our findings suggest that researchers should establish classification models to support the identification of erosion symptoms and pay more attention to the evolution of erosion symptoms. Practitioners should manage the ignored erosion symptoms with their evolution and continuously measure the impact of the erosion symptoms during the development life cycle. Besides, practitioners should be vigilant about the potential risk of architecture erosion to avoid erosion symptoms transferring to technical debt in a long run. To summarize, code review as a code inspection activity can help to find out and remove potential architecture erosion symptoms and to some extent prevent architecture erosion.


As a next step, we plan to empirically evaluate tools that can identify architecture erosion from source code and compare the results with associated artifacts (e.g., commits, issues) containing architecture erosion symptoms, as well as to investigate which symptoms cannot be identified by the existing tools.



\balance
\bibliographystyle{IEEEtran}
\bibliography{ref}
\end{document}